\documentclass[cits]{PoS}

\def\apj{ApJ}%
\def\apss{Ap\&SS}%
\def\aap{A\&A}%
\def\mnras{MNRAS}%
\title{Spectral states in Be/X-ray pulsars}

\ShortTitle{Spectral states in Be/X-ray pulsars}

	\author{\speaker{Elisa Nespoli}\\
         - Observatorio Astron\'omico de la Universidad de Valencia, Calle Catedr\'atico Jose Beltran, 2, 46980 Paterna (Valencia), Spain\\	
       - European Space Astronomy Centre (ESA/ESAC), Science Operations Department, Villanueva de la Ca\~nada (Madrid), Spain\\
       E-mail: \email{elisa.nespoli@uv.es}}

\author{Pablo Reig\\
        - IESL, Foundation for Research and Technology-Hellas, 71110, Heraklion, Greece. \\
        - Institute of Theoretical \& Computational Physics, University of Crete, PO Box 2208, GR-710 03, Heraklion, Crete, Greece.\\
        E-mail: \email{pau@physics.uoc.gr}}

\abstract{In the last quarter of a century, a unified characterization of the spectral evolution of low-mass X-ray binaries, both containing a neutron star and a black hole, was possible. 
In this context, the notion of source states characterizing the X-ray emission from black-hole binaries and neutron-star low-mass X-ray binaries revealed to be a very useful tool to disentangle the complex spectral and aperiodic phenomenology displayed by those classes of accreting objects. Be/X-ray binaries constitute another major class of transient accreting binaries, for which very little work has been done on the correlated timing and spectral variability. Especially, no definition of source states exists for this class, in spite of their highly variable X-ray emission. When active, Be/X-ray binaries are among the brightest objects in the X-ray sky and are characterized by dramatic variability in brightness on timescales ranging from seconds to years. It is then worth it to ask whether a definition of spectral states is possible for these systems. 

In this work, we try to address such a question, investigating whether accreting X-ray pulsars display source states and characterizing those states through their spectral properties. Our results show that Be/X-ray pulsars trace two different branches in their hardness-intensity diagram: the \emph{horizontal branch},  a low-intensity state, and the \emph{diagonal branch}, a high-intensity state that only appears when the X-ray luminosity exceeds a critical limit. We propose that the two branches are the phenomenological signature of two different accretion modes -- in agreement with recently proposed models -- depending on whether the luminosity of the source is above or below a critical value. 
}

\FullConference{ An INTEGRAL view of the high-energy sky (the first 10 years) - 9th INTEGRAL Workshop and celebration of the 10th anniversary of the launch\\
		 15-19 October 2012\\
		 Bibliotheque Nationale de France, Paris, France}

\begin{document}

\section{Introduction}

In the last twenty-five years, a unified description of the spectral evolution of  low-mass X-ray binaries (LMXB) and black-hole binaries (BHB) was made possible \cite{has89,hom05,dun10,kle08}. In such systematic studies, color-color diagrams (CDs) and hardness-intensity diagrams (HIDs) emerged as powerful tools to characterize those systems, since it was observed that, both for persistent and transient systems, the source emission evolves smoothly along the diagrams, describing a pattern. Similarly to color-magnitude diagrams in the optical (although with much faster time scales), CDs and HIDs permitted to unveil and address questions that could not be managed otherwise, by introducing the crucial notion of \emph{spectral states}. A state is defined by the specific and well-defined position that the sources occupy in their CDs/HIDs, and is associated to the appearance (or disappearance) of a characteristic spectral or timing component in the corresponding energy spectra and power spectral density.

Within high-mass X-ray binaries (HMXBs), Be/X-ray binaries (Be/XRBs) represent the major subclass, and one of the most extreme case of X-ray variability on several time scales. These systems consist in a strongly magnetized neutron star orbiting a massive Be star. The observed X-ray emission is powered by large amount of matter accreted onto the compact object, during two different classes of events:
\vspace{-0.1cm}
\begin{itemize}
\item[-] type-I (or normal) outbursts, lasting a few days, with luminosities $\sim$10$^{36}$-- 10$^{37}$ erg s$^{-1}$, occur in series separated by the orbital period, generally close to the time of periastron passage of the neutron star;
\item[-]type-II (or giant) outbursts, rarer and unpredictable, last several weeks, showing  2--3 orders of magnitude variations in X-ray intensity ($L_\mathrm{X} > 10^{37}$ erg s$^{-1}$). 
\end{itemize} 

This work focuses on the second type of X-ray outbursts, with the purpose of investigating the spectral variability in Be/XRBs as a class, and the final objective of defining and characterizing spectral states. A first attempt to define spectral states through color analysis was carried out by \cite{rei06, rei08}. The present work builds up on those results, extends them to include all the Be/XRBs detected by RXTE during a type-II outburst, presents a finer sampling resolution in spectral analysis, and provides an interpretation of the identified states within the recently proposed scenario that describes accretion in Be/XRBs.

\section{Analysis}

We analyzed archived data from the \emph{Rossi X-ray Timing Explorer} (RXTE). The hardness, or soft color (SC), was obtained directly from the Proportional Counter Array (PCA) count rates in the  7-10 keV / 4-7 keV energy bands. Energy spectra were extracted, one spectrum per pointing, for the PCA and the High-energy Timing Experiment (HEXTE), and simultaneously fitted. The lack of adequate theoretical continuum models for accreting neutron stars implies the use of empirical models to fit the observations. To compare the results from all sources consistently, we used the same spectral components for all the system analyzed. The best-fit model consisted in a combination of photoelectric absorption and a power-law with high-energy exponential cutoff to fit the continuum. The discrete components were accounted for with an emission Gaussian line profile, which fits the iron Fe K$\alpha$ fluorescence, and Lorentzian absorption profiles, which fitted cyclotron resonance scattering features (CRSF). For three sources, 4U 0115+63, XTE J0658--073 and 1A 0535+262, additional residuals around 7--10 keV required the inclusion of a broad Gaussian emission line, a feature known as the ``bump'' model \cite{fer09}. 
Table~\ref{xobs} shows the journal of observations, including the maximum X-ray luminosity reached at the outburst peak, and a measure of the central energy of the CRSF, when present, and the average $\chi_\mathrm{red}^{2}$ for the spectral fitting.

\begin{table}
\begin{center}
\resizebox{10cm}{!}{
\begin{tabular}{@{~~}l@{~~}c@{~~}c@{~~}c@{~~}c|@{~~}c|@{~~}c}
\hline\hline
Source &Time range &Num. &Exp. & Max. $L_\mathrm{X}$ & $E_\mathrm{cyc}$ & $\chi_\mathrm{red}^{2}$ \\
name  & (MJD) &obs &time (ks)  & (erg s$^{-1}$) & (keV) & \\
\hline
4U 0115+63  &53254.1--53304.6 &36 &96.4  & 1.4$\times10^{38}$  & 12 & $1.5\pm0.4$  \\
KS 1947+300  &51874.2--52078.0 &86 &144.9 & 7.1$\times10^{37}$ & --  &  $1.1\pm0.4$\\
EXO 2030+375  &53914.9--54069.9 &154 &140.1 & 1.6$\times10^{38}$ & 11 &  $1.1\pm0.1$ \\
V 0332+53  &53340.3--53376.6 &80 & 174.5  & 3.4$\times10^{38}$ & 25 & $1.6\pm0.5$  \\
1A 1118--616  &54838.3--54865.0 &26 & 10.2  & 2.9$\times10^{37}$ & 60  & $1.2\pm0.2$  \\
XTE J0658--073  &52932.7--53056.4 & 76 &201.9  & 3.6$\times10^{37}$ & 33 & $1.1\pm0.3$  \\
Swift J1656.6--5156&53723.9--54269.9 &260 & 313.0 & 5.2$\times10^{37}$  & -- & $0.9\pm0.2$  \\
1A 0535+262 &55169.4--55210.5 &27 &126.7 & 6.8$\times10^{37}$ & 46 & $0.7\pm0.2$    \\
GRO J1008--57 &54426.0--54457.2 &20 &27.5  & 2.0$\times10^{37}$ & -- &  $0.8\pm0.2$\\
\hline
\vspace{-1cm}
\end{tabular}}
\end{center}
\caption{RXTE observations.}
\label{xobs}
\end{table}

\section{Evidence for source states in Be/XRBs}

In this section we present our results, which unveil the existence of spectral states in X-ray pulsars from both model-independent color analysis and spectral analysis.

\subsection{Hardness-intensity diagrams}

\begin{figure} 
\centering
\includegraphics[width=.8\textwidth]{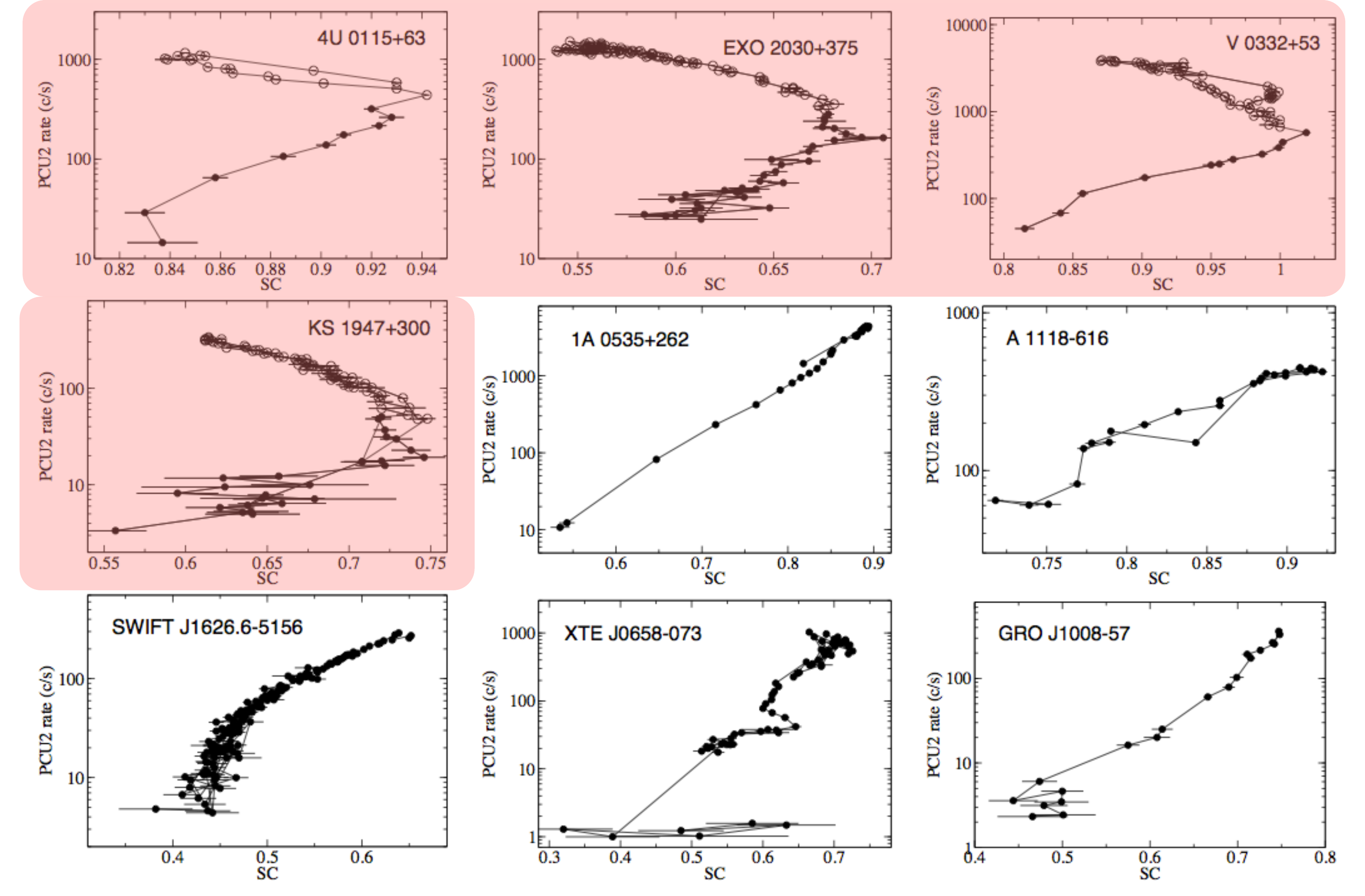}
\caption{Hardness-intensity diagrams for the nine systems studied in this work.}
\label{hid} 
\end{figure}

For all the sources analyzed in this work., Fig.~\ref{hid} shows the corresponding HID, a plot of the 3--30 keV band count rate as a function of hardness, one point per RXTE observation. 
In the figure, a color shade marks the four sources which display two branches in their HID, with a clear transition between them: 4U 0115+63, EXO 2030+375, V 0332+53, and KS 1947+300. These correspond to the brightest sources (see Table 1), although the luminosity does not seem to be the only parameter triggering the state transition.
For consistency with \cite{rei06, rei08}, we will keep the terminology \emph{horizontal branch} (HB) and \emph{diagonal branch} (DB) to designate the low-luminosity branch and the high-luminosity branch, respectively, although in the present work, as we are plotting the HID using a logarithmic representation, both branches appear as diagonal in the diagram.

Due to the unpredictable nature of type-II outbursts, it is usually difficult to have a proper coverage of the very first phase of the rise. Nevertheless, for one system, KS 1947+300, data were collected during the full duration of the outburst. We shall thus take KS 1947+300 as a paradigm of the objects displaying the two branches, and use it as an example to describe the pattern described by the source during the outburst. 

\begin{figure}[h!]
\centering
\includegraphics[width=.9\textwidth]{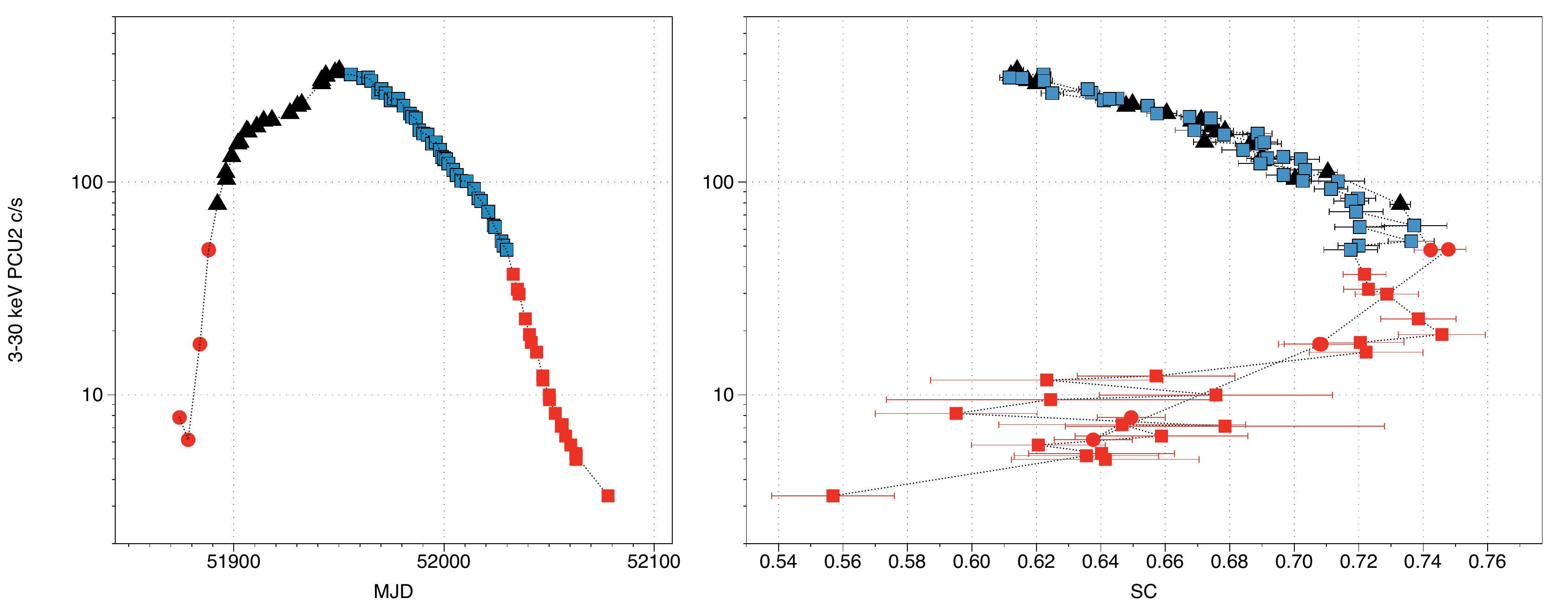}
\caption{Light curve \emph{(left panel)} and HID \emph{(right panel)} for KS 1947+300}
\label{kshid} 
\end{figure}

Figure~\ref{kshid} shows the outburst profile and HID for KS 1947+300, where we marked with different colors/symbols the different phases of the outburst. When the outburst begins, as the flux increases, the SC increases, \emph{i.e.} the emission from the source becomes harder and harder, and the system is moving rightward on the HID, depicting the HB (red circles). When the luminosity reaches a critical value, the source undergoes the transition, and makes a sudden turn in the HID, entering the DB (black triangles). At this stage, as the flux increases, the SC decreases, \emph{i.e.} X-ray emission becomes softer and softer, and the system moves leftward on the HID, until the peak luminosity is reached. Then, as flux starts decreasing, the hardness starts increasing again, and the system moves back on the DB, tracing the same pattern as during the rise (blue squares).

The transition occurs again at about the same critical luminosity as in the rise, when the source leaves the DB to enter the HB, and the hardness decreases as the flux decreases (red squares), back to the starting point in the HID. The two branches are drawn in exactly the same way during the rise and the decay, depending only on the luminosity and not on the outburst rise/decay phase, \emph{i.e.}, no hysteresis is observed.

\subsection{Spectral analysis}

\begin{figure}[h!]
\centering
\includegraphics[width=.75\textwidth]{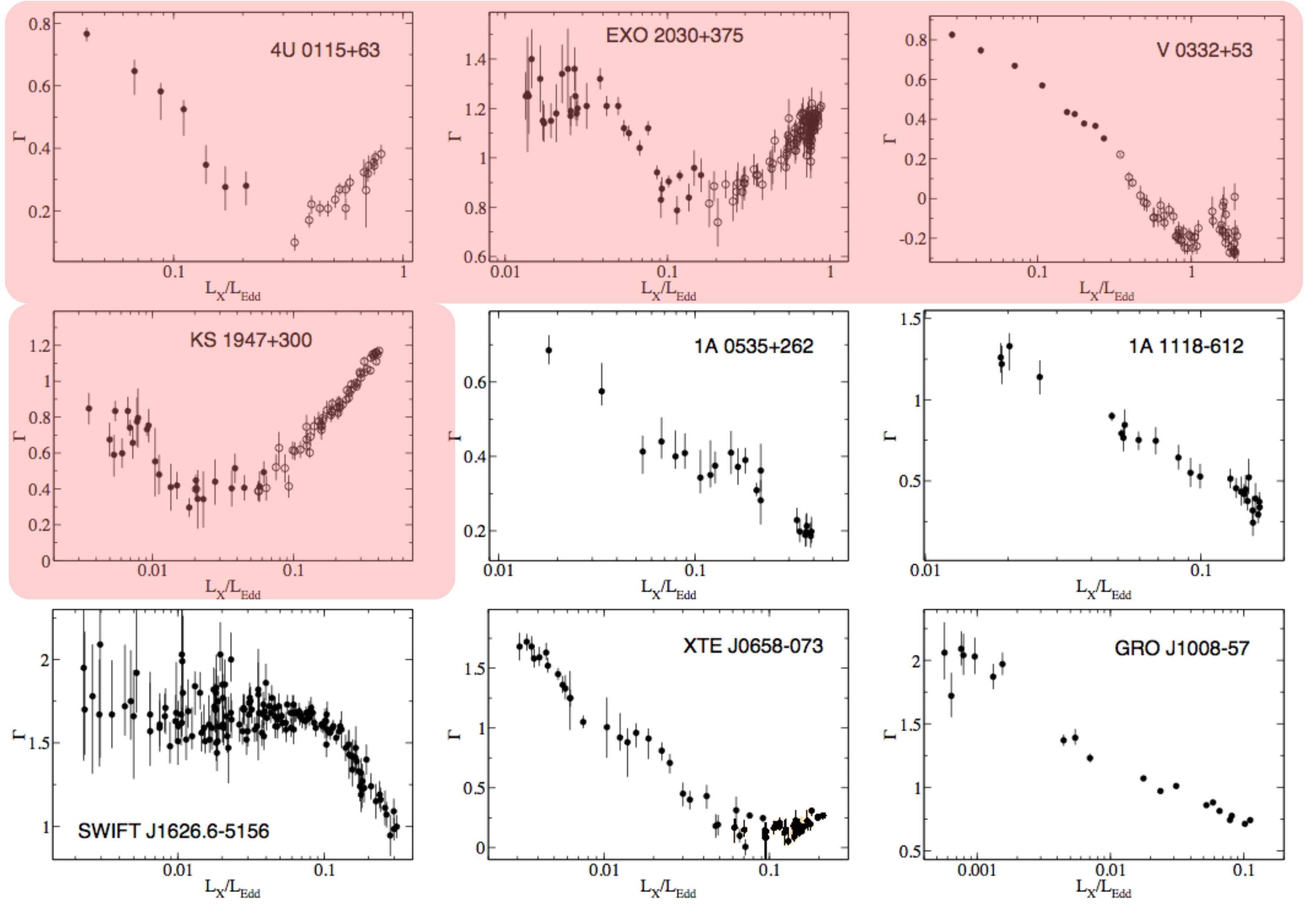}
\caption{Photon index as a function of X-ray luminosity.}
\label{gamma} 
\end{figure}

The main parameter representing the spectral continuum in the spectral continuum in X-ray pulsars in the classical X-ray range  is the photon index $\Gamma$. Figure~\ref{gamma} shows the relation between $\Gamma$ and luminosity, where $L_\mathrm{Edd}=1.7\times10^{38}$ erg s$^{-1}$ is the Eddington luminosity for a neutron star.

The same four systems that show two branches in their HID, also display a bimodal relation of $\Gamma$ with luminosity, passing from a negative correlation to a positive correlation, after a critical luminosity is reached, confirming the existence of two luminosity spectral states. Instead, only one state is shown by the remaining systems. In Fig.~\ref{gamma}, we marked with filled/open circles the pointings corresponding to the HB/DB respectively, as defined by the HID. One can see that there is fully agreement between results from color and spectral analysis. Although correlations between the photon index and luminosity have been reported in the past, we report for the first time a sudden change in the correlation sign, depending on the position of the source in its HID.

\section{Spectral states as a signature of a bimodal accretion}

Spectral variability in X-ray pulsars is thought to arise close to the neutron star surface by means of bulk and thermal Comptonization in an accretion column. This would form in correspondence of one or both polar caps, where the accretion flow is channelled by the neutron star magnetosphere \cite{bec07}. A detailed description of the interaction between the highly magnetized plasma surrounding the pulsar and the accretion flow is still lacking, although recently, an attempt to phenomenologically model the vicinity of the neutron star surface was able to describe the bimodal correlation observed between the CRSF energy and luminosity \cite{bec12}. Those authors invoked two different regimes of accretion depending on a critical luminosity, $L_\mathrm{crit}$. Below (but close to) $L_\mathrm{crit}$, the deceleration of the accreting flow at the neutron star surface would be mainly due to Coulomb interaction, while above that value, the main decelerating factor would be radiation pressure. In this work we found that X-ray pulsars display two spectral states, depending on the luminosity they reach during an outburst. Only the brightest sources enter the DB. We measured the transitional luminosity in the range $\sim$ 1--4 $\times 10^{37}$ erg s$^{-1}$, depending on the source. We propose that the transitional luminosity correspond to the critical luminosity in \cite{bec12}, and that the two branches we observe are a signature of the two accretion regimes, being the HB the sub-critical state, and the DB the super-critical one. 
In \cite{bec12}, the authors propose that the critical luminosity is proportional to the neutron star magnetic field intensity, or equivalently, to the energy of the cyclotron line. This is consistent with what we found. In fact, KS 1947+300 and 1A 0535+262 reached almost the same luminosity (see Table~\ref{xobs}), but only the first system enters the DB. This behavior would be conveniently explained by the dependence of $L_\mathrm{crit}$ on the magnetic field, which for 1A 0535+262 is much higher and would thus prevent the source to undergo the transition.

\section{Conclusions}

We have performed color and spectral analysis of nine Be/XRBs during a giant outburst. The evolution through the hardness-intensity diagram suggests that Be/XRBs undergo state transitions, displaying, if a critical luminosity is reached, two different source states. These results are corroborated by spectral analysis. The two states we observe are a plausible evidence of the two accretion regimes proposed by \cite{bec12}. For a more detailed analysis of the results presented here the reader is referred to \cite{rei12}

\end{document}